\documentclass[RNAAS]{aastex63}

\usepackage{hyperref}

\shorttitle{Merger fraction evolution 
in CLAUDS+HSC-SSP}
\shortauthors{Thibert et al.}
\graphicspath{{./}{figures/}}

\begin{document}

\title{{Evolution of the galaxy merger fraction in the CLAUDS+HSC-SSP deep fields}}

\correspondingauthor{Marcin  Sawicki}
\email{marcin.sawicki@smu.ca}

\author{Nathalie Thibert}
\affiliation{Institute for Computational Astrophysics and Department of Astronomy and Physics, Saint Mary's University, Halifax, NS, B3H 3C3, Canada}

\author{Marcin Sawicki}
\affiliation{Institute for Computational Astrophysics and Department of Astronomy and Physics, Saint Mary's University, Halifax, NS, B3H 3C3, Canada}

\author{Andy Goulding}
\affiliation{
Department of Astrophysical Sciences, Princeton University, Princeton, NJ 08544-1001, USA}

\author{St\'ephane Arnouts}
\affiliation{
CNRS, LAM - Laboratoire d'Astrophysique de Marseille, Aix Marseille Université, 38 rue F. Joliot-Curie, F-13388 Marseille, France
}

\author{Jean Coupon}
\affiliation{
Astronomy Department, University of Geneva, Chemin d’Ecogia 16, CH-1290 Versoix, Switzerland
}

\author{Stephen Gwyn}
\affiliation{
NRC Herzberg Astronomy and Astrophysics, 5071 West Saanich Road, Victoria, BC, V9E 2E7, Canada
}



\begin{abstract}
We estimate the evolution of the galaxy-galaxy merger fraction for $M_\star>10^{10.5}M_\odot$ galaxies over $0.25<z<1$ in the $\sim$18.6~deg$^2$ deep CLAUDS+HSC-SSP surveys.   We do this by training a Random Forest Classifier to identify merger candidates from a host of parametric morphological features, and then visually follow-up likely merger candidates to reach a high-purity, high-completeness merger sample. 
Correcting for redshift-dependent detection bias, we find that the merger fraction at $z=0$  is 1.0$\pm$0.2\%, that the merger fraction evolves as $(1+z)^{2.3 \pm 0.4}$, and that a typical massive galaxy has undergone $\sim$0.3 major mergers since $z=1$.  This pilot study illustrates the power of very deep ground-based imaging surveys combined with machine learning to detect and study mergers through the presence of faint, low surface brightness merger features out to at least $z\sim1$. 
\end{abstract}




\section{Introduction} \label{sec:introduction
}
Galaxy-galaxy mergers play an important role in  the hierarchical structure formation paradigm as they provide avenues for galaxies to grow their stellar masses and transform their morphologies (e.g., \citealt{Bundy2009, Xu2012, Conselice2014}). Though rare today, merging of similar-mass galaxies was common in the past \citep[e.g.,][]{LeFevre2000, Patton2002, Bridge2010, Conselice2014, Sawicki2020}. To further characterize the role of mergers over cosmic time requires samples sufficiently large to study the dependence of merging on redshift, environment, and intrinsic galactic properties such as mass, star formation rate, or AGN activity. 

Distant merging galaxies can be identified as close pairs, likely to merge through dynamical friction \citep[e.g.,][]{Patton2002, Bundy2009, Mundy2017}, but this approach requires statistical background corrections or expensive spectroscopy. An alternative is to find galaxies with disturbed morphologies marking ongoing mergers \citep[e.g.,][]{LeFevre2000, Bridge2010}.  Identification of such morphological signatures can be done visually, but this is tedious for large samples and not easily reproducible. Consequently, automated approaches to morphological merger identification are attractive \citep[e.g.,][]{Conselice2008, Lotz2008, Freeman2013,Ackermann2018,Goulding2018}. Here we summarize the results of our pilot study (\citealt{Thibert2018}\footnote{Full text:  \url{https://library2.smu.ca/handle/01/27980}}) that uses morphological features and a Random Forest Classifier (RFC) to pre-select merger candidates in the combined Hyper Suprime-Cam Strategic Survey Program  \citep[HSC-SSP;][]{Aihara2018} and CFHT Large Area U-band Deep Survey  \citep[CLAUDS;][]{Sawicki2019} dataset.

\section{Data} \label{sec:data}

\begin{figure}[ht!]
\begin{center}
\includegraphics[scale=0.73,angle=0]{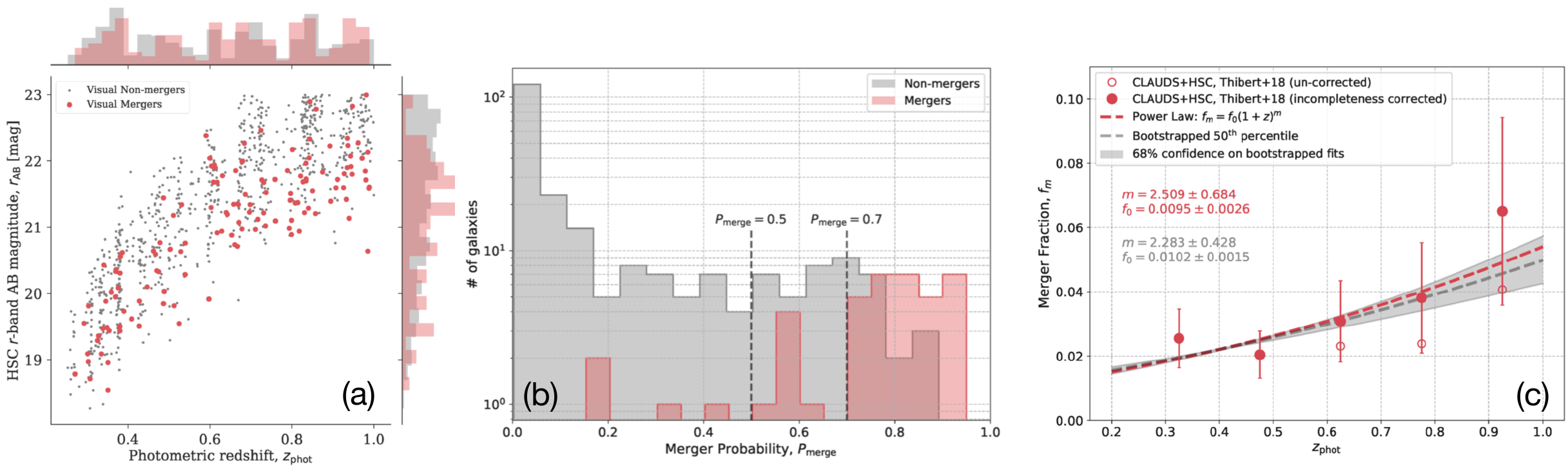}
\caption{
(a) Our training/testing sample; red points mark galaxies visually classified as mergers. The $M_\star > 10^{10.5} M_\odot$ cut produces the upper envelope and only at the highest redshift the $r_{AB}<23.0$ cut starts becoming important. 
(b) Distribution of merger probabilities reported by the RFC for the testing sample;  red are visually-classified mergers, gray are non-mergers. 
(c) Merger fraction evolution. Open points denote the raw merger fraction, while closed points show the incompleteness-corrected values. The red dashed line is the power-law fit to the incompleteness-corrected points. The gray dashed line and shaded region denote the 50th percentile and 68\% confidence interval from  bootstrapped fits to the (incompleteness-corrected) measurements.
\label{TheFigure}
}
\end{center}
\end{figure}

We use the  photometric redshifts and stellar masses ($M_*$) estimated from very deep $u/u^*$+$grizy$ photometry of the overlapping 18.60~deg$^2$ of the CLAUDS and HSC-SSP Deep/UltraDeep regions.  Our $z_{phot}$ are described in \citet{Sawicki2019} while $M_*$ estimates are similar to those of \cite{Moutard2020}. We match this catalog to the $r$-band images from an internal HSC-SSP data release intermediate in depth between PDR1 and PDR2 \citep{PDR1, PDR2}. Visual inspection suggests that $r_{AB}<23.0$ is sufficient for morphological feature detection. Together with a  $M_\star < 10^{10.5}M_\odot$ stellar mass cut, this gives  60,957 galaxies with $0.25<z<1$. 

\section{Morphological Merger Detection} \label{sec:method}

We measure the following $r$-band  morphological features (see \citealt{Goulding2018}) for our galaxies:
Sersic index ($n_{Sersic}$), 
concentration ($C$), 
asymmetry ($A$), 
residual asymmetry ($A_{resid}$), 
residual flux fraction ($RFF$), 
smoothness ($S$), 
residual smoothness ($S_{resid}$), 
Gini coefficient ($G$), 
moment of light ($M_{20}$), 
and Petrosian radius ($R_{Petro}$). 
Together with star-formation probability ($P_{SF}$, from color-color classification), photo-$z$, $r_{AB}$, and $M_\star$, these parameters are used in our Random Forest Classifier (RFC). 

We then visually inspect the images of a random, uniform sub-sample of 918 galaxies, to each assigning a merger/non-merger label (136 mergers, 782 non-mergers). This constitutes our training/testing sample (Figure~\ref{TheFigure}(a)).  We train the RFC using 70\% (642 objects) of our visually-classified sample, and hold back 30\% (276) for testing. To balance classes in training, we randomly sub-sample from the non-mergers while up-boosting the merger population by constructing artificial objects based on the features of the identified mergers. The RFC finds that the features in order of decreasing importance are $RFF$, $M_{20}$, $S_{resid}$, $n_{Sersic}$, $C$, $A$, $P_{SF}$, $A_{resid}$, $R_{petro}$, $S$, $G$, $z_{phot}$, $r_{AB}$, and $M_\star$.

For each galaxy, the RFC produces an ensemble of binary classifications, and we take the average of these as the merger probability, $P_{merge}$. Testing the RFC output shows that $P_{merge}>0.7$ identifies most of the visually-classified mergers, although with a significant fraction of non-merger contaminants (Figure~\ref{TheFigure}(b)). We thus visually inspect the 7,550 objects (13\% of the sample) with $P_{merge}>0.7$, yielding 1,576 secure mergers. 

\section{Merger Fraction Evolution} \label{sec:results}

Dividing the secure merger numbers by the total galaxy sample gives the {\emph{raw, directly observed}} merger fraction  (open points in Figure~\ref{TheFigure}(c)). 
However, morphological merger features become harder to detect with increasing redshift, necessitating a redshift-dependent incompleteness correction.  We estimate this correction by selecting eight low-redshift mergers from among our images, artificially dimming and spatially resampling them to simulate their appearance at higher redshifts, inserting them into blank-sky regions within our images, and then running the resulting images through our RFC+visual-followup procedure employed previously. The redetection rate gives an estimate of the redshift-dependent incompleteness correction, and the resulting incompleteness-corrected values are shown with filled points in Figure~\ref{TheFigure}(c). 
Although the size of the uncertainties, particularly at higher redshifts, is now dominated by the Poisson noise in our incompleteness correction measurement, we measure a merger fraction evolution,  $f_{merge}(z) = (0.010 \pm 0.002)\times(1+z)^{2.3\pm0.4}$ (Figure 1(c), gray curve), consistent with previous studies (see Table~5.4 of \citealt{Thibert2018} for a compilation).  Integrating these values suggests that a typical massive galaxy has undergone $\sim$0.3 major mergers since $z=1$.

\section{Discussion and conclusions} \label{sec:discussion}

Deep ground-based imaging is more sensitive to low surface brightness merger signatures than space-based imaging from HST. Our pilot study points to machine learning as a useful approach to morphologically detecting mergers out to at least $z\sim1$ in deep HSC-SSP imaging and similarly-deep datasets (e.g., LSST). In the near future, we will perform ML-based merger detection on the deeper  HSC-SSP PDR3 images to study merger fractions as function of redshift, environment, and galaxy evolutionary state.

\bibliography{mergers}{}

\begin{thebibliography}{}
\expandafter\ifx\csname natexlab\endcsname\relax\def\natexlab#1{#1}\fi
\providecommand{\url}[1]{\href{#1}{#1}}
\providecommand{\dodoi}[1]{doi:~\href{http://doi.org/#1}{\nolinkurl{#1}}}
\providecommand{\doeprint}[1]{\href{http://ascl.net/#1}{\nolinkurl{http://ascl.net/#1}}}
\providecommand{\doarXiv}[1]{\href{https://arxiv.org/abs/#1}{\nolinkurl{https://arxiv.org/abs/#1}}}

\bibitem[{Ackermann {et~al.}(2018)Ackermann, Schawinski, Zhang, Weigel, \&
  Turp}]{Ackermann2018}
Ackermann, S., Schawinski, K., Zhang, C., Weigel, A.~K., \& Turp, M.~D. 2018,
  MNRAS, 479, 415, \dodoi{10.1093/mnras/sty1398}

\bibitem[{Aihara {et~al.}(2018{\natexlab{a}})Aihara, Arimoto, Armstrong,
  Arnouts, Bahcall, Bickerton, Bosch, Bundy, Capak, Chan, \&
  et~al.}]{Aihara2018}
Aihara, H., Arimoto, N., Armstrong, R., {et~al.} 2018{\natexlab{a}}, PASJ, 70,
  S4, \dodoi{10.1093/pasj/psx066}

\bibitem[{Aihara {et~al.}(2018{\natexlab{b}})Aihara, Armstrong, Bickerton,
  Bosch, Coupon, Furusawa, Hayashi, Ikeda, Kamata, Karoji, \& et~al.}]{PDR1}
Aihara, H., Armstrong, R., Bickerton, S., {et~al.} 2018{\natexlab{b}}, PASJ,
  70, S8, \dodoi{10.1093/pasj/psx081}

\bibitem[{Aihara {et~al.}(2019)Aihara, AlSayyad, Ando, Armstrong, Bosch, Egami,
  Furusawa, Furusawa, Goulding, Harikane, \& et~al.}]{PDR2}
Aihara, H., AlSayyad, Y., Ando, M., {et~al.} 2019, PASJ, 71, 114,
  \dodoi{10.1093/pasj/psz103}

\bibitem[{Bridge {et~al.}(2010)Bridge, Carlberg, \& Sullivan}]{Bridge2010}
Bridge, C.~R., Carlberg, R.~G., \& Sullivan, M. 2010, ApJ, 709, 1067,
  \dodoi{10.1088/0004-637X/709/2/1067}

\bibitem[{Bundy {et~al.}(2009)Bundy, Fukugita, Ellis, Targett, Belli, \&
  Kodama}]{Bundy2009}
Bundy, K., Fukugita, M., Ellis, R.~S., {et~al.} 2009, ApJ, 697, 1369,
  \dodoi{10.1088/0004-637X/697/2/1369}

\bibitem[{Conselice(2014)}]{Conselice2014}
Conselice, C.~J. 2014, ARAA, 52, 291,
  \dodoi{10.1146/annurev-astro-081913-040037}

\bibitem[{Conselice {et~al.}(2008)Conselice, Rajgor, \& Myers}]{Conselice2008}
Conselice, C.~J., Rajgor, S., \& Myers, R. 2008, MNRAS, 386, 909,
  \dodoi{10.1111/j.1365-2966.2008.13069.x}

\bibitem[{Freeman {et~al.}(2013)Freeman, Izbicki, Lee, Newman, Conselice,
  Koekemoer, Lotz, \& Mozena}]{Freeman2013}
Freeman, P.~E., Izbicki, R., Lee, A.~B., {et~al.} 2013, MNRAS, 434, 282,
  \dodoi{10.1093/mnras/stt1016}

\bibitem[{Goulding {et~al.}(2018)Goulding, Greene, Bezanson, Greco, Johnson,
  Leauthaud, Matsuoka, Medezinski, \& Price-Whelan}]{Goulding2018}
Goulding, A.~D., Greene, J.~E., Bezanson, R., {et~al.} 2018, PASJ, 70,
  \dodoi{10.1093/pasj/psx135}

\bibitem[{Le~Fevre {et~al.}(2000)Le~Fevre, Abraham, Lilly, Ellis, Brinchmann,
  Schade, Tresse, Colless, Crampton, Glazebrook, \& et~al.}]{LeFevre2000}
Le~Fevre, O., Abraham, R., Lilly, S.~J., {et~al.} 2000, MNRAS, 311, 565–575,
  \dodoi{10.1046/j.1365-8711.2000.03083.x}

\bibitem[{Lotz {et~al.}(2008)Lotz, Davis, Faber, Guhathakurta, Gwyn, Huang,
  Koo, Le~Floc’h, Lin, Newman, \& et~al.}]{Lotz2008}
Lotz, J.~M., Davis, M., Faber, S.~M., {et~al.} 2008, ApJ, 672, 177,
  \dodoi{10.1086/523659}

\bibitem[{Moutard {et~al.}(2020)Moutard, Sawicki, Arnouts, Golob, Coupon,
  Ilbert, Yang, \& Gwyn}]{Moutard2020}
Moutard, T., Sawicki, M., Arnouts, S., {et~al.} 2020, MNRAS, 494, 1894,
  \dodoi{10.1093/mnras/staa706}

\bibitem[{Mundy {et~al.}(2017)Mundy, Conselice, Duncan, Almaini, Häußler, \&
  Hartley}]{Mundy2017}
Mundy, C.~J., Conselice, C.~J., Duncan, K.~J., {et~al.} 2017, MNRAS, 470, 3507,
  \dodoi{10.1093/mnras/stx1238}

\bibitem[{Patton {et~al.}(2002)Patton, Pritchet, Carlberg, Marzke, Yee, Hall,
  Lin, Morris, Sawicki, Shepherd, \& et~al.}]{Patton2002}
Patton, D.~R., Pritchet, C.~J., Carlberg, R.~G., {et~al.} 2002, ApJ, 565, 208,
  \dodoi{10.1086/324543}

\bibitem[{Sawicki {et~al.}(2020)Sawicki, Arcila-Osejo, Golob, Moutard, Arnouts,
  \& Cheema}]{Sawicki2020}
Sawicki, M., Arcila-Osejo, L., Golob, A., {et~al.} 2020, MNRAS, 494, 1366,
  \dodoi{10.1093/mnras/staa779}

\bibitem[{Sawicki {et~al.}(2019)Sawicki, Arnouts, Huang, Coupon, Golob, Gwyn,
  Foucaud, Moutard, Iwata, Liu, \& et~al.}]{Sawicki2019}
Sawicki, M., Arnouts, S., Huang, J., {et~al.} 2019, MNRAS, 489, 5202,
  \dodoi{10.1093/mnras/stz2522}

\bibitem[{Thibert(2018)}]{Thibert2018}
Thibert, N. C.~M. 2018, MSc Thesis, Saint Mary’s University.
\newblock \url{http://library2.smu.ca/handle/01/27980}

\bibitem[{Xu {et~al.}(2012)Xu, Zhao, Scoville, Capak, Drory, \& Gao}]{Xu2012}
Xu, C.~K., Zhao, Y., Scoville, N., {et~al.} 2012, ApJ, 747, 85,
  \dodoi{10.1088/0004-637X/747/2/85}

\end{thebibliography}
\bibliographystyle{aasjournal}



\end{document}